\documentclass[12pt]{article}
\usepackage{cite}
\usepackage{graphicx}
\usepackage{ amssymb } 
\usepackage{upgreek}       
\usepackage{textcomp} 
\pdfoutput=1

\begin{document}

\newcommand{\titulo}{Calculations of {Some} Doping Nanostructuration{s}\\ and Pattern{s Improving} the Functionality\\ of High-Temperature Superconductors\\ for Bolometer Device Applications}

\newcommand{\autor}{
{J.C. Verde}$^a$, {A.S.} {Viz} $^{a,b}$, {M.M.} Botana $^{a,b}$, {C.} Montero-Orille $^{b,c}$ and {M.V.} Ramallo $^{a,b}$
}

\newcommand{\direccion}{
$^{a}$  Quantum Materials and Photonics Research Group (QMatterPhotonics), Department of Particle Physics, University of Santiago de Compostela,\\ ES-15782 Santiago de Compostela, Spain\\ \mbox{}\\

$^{b}$  Strategic Grouping in Materials AeMAT, University of Santiago de Compostela, ES-15782 Santiago de Compostela, Spain\\ \mbox{}\\

$^{c}$  Quantum Materials and Photonics Research Group (QMatterPhotonics), Department of Applied Physics, University of Santiago de Compostela,\\ ES-15782 Santiago de Compostela, Spain}

\begin{center}
  \Large\bf
\titulo\\  \end{center}\mbox{}\vspace{-2cm}\\ 

\begin{center}\normalsize\autor\end{center} \mbox{}\vspace{-1cm}\\

\begin{center}\normalsize\it\direccion\end{center}

\newcommand{\eg}{e.g.}
\newcommand{\ie}{i.e.}
\newcommand{\vs}{vs.}
\newcommand{\etc}{etc.}
\newcommand{\etal}{\textit{et al.}}
\newcommand{\Tc}{\mbox{$T_c$}}
\newcommand{\Tback}{\mbox{$T_c^{\rm back}$}}
\newcommand{\Tdot}{\mbox{$T_c^{\rm dot}$}}
\newcommand{\gsim}{\stackrel{>}{_\sim}}
\newcommand{\dd}{\mbox{d}}

\newcommand{\thermal}{\cite{HTSbol9,HTSbol4,HTSbol1,HTSbol5,HTSbol11,Oktem,HTSbol6,HTSbol7,Tkachenko,Kaiser}}
\newcommand{\mmwave}{\cite{HTSbol9,HTSbol4,HTSbol1,HTSbol5,HTSbol11,Oktem,HTSbol7,Tkachenko, Kaiser,HTSbol8,Nivelle,HTSbol12}}
\newcommand{\meander}{\cite{HTSbol9,HTSbol4,HTSbol1,HTSbol5,HTSbol11,Oktem,HTSbol7,Tkachenko, Kaiser,HTSbol8,Nivelle}}
\newcommand{\spacebased}{\cite{HTSbol8,Nivelle}}
\newcommand{\labfi}{\cite{Nivelle}}
\newcommand{\bolometros}{\cite{HTSbol9,HTSbol8,HTSbol4,HTSbol1,HTSbol5,HTSbol11,HTSbol12,Oktem ,HTSbol3,HTSbol6,HTSbol7,Khrebtov,Tkachenko,Nivelle,Kaiser,Frenkel, Nature2,Harry,Irwin,Abdel}}

\newcommand{\YBCO}{YBCO}


\mbox{}\vskip0.0cm{\bf Abstract: } \quad  We calculate the effects of doping nanostructuration and the patterning of thin films of high-temperature superconductors (HTS) with the aim of optimizing their functionality as sensing materials for resistive transition-edge bolometer devices (TES).
{We focus, in particular, on spatial variations of the carrier doping into the CuO$_2$ layers due to oxygen off-stoichiometry, (that~induce, in turn, critical temperature variations) and explore following two major cases of such structurations:}
First,~the~random nanoscale disorder intrinsically associated to doping levels that do not maximize the superconducting critical temperature; our studies suggest that this first simple structuration already improves some of the bolometric operational parameters with respect to the conventional, nonstructured HTS materials used until now. Secondly, we consider the imposition of regular arrangements of zones with different nominal doping levels (patterning); we find that such regular patterns may improve the bolometer performance even further. We find one design that improves, with respect to nonstructured HTS materials, both the saturation power and the operating temperature width by more than one order of magnitude. It also almost doubles the response of the sensor to radiation.

\vfill

\mbox{}\hfill{\footnotesize {\tt mv.ramallo@usc.es}}
\thispagestyle{empty}

\newpage
\setlength{\baselineskip}{18pt}
\



\section{Introduction}

Bolometers are radiation sensors that detect incident energy via the increase of the temperature $T$ caused by the absorption of incoming photons~\cite{HTSbol9,HTSbol8,HTSbol4,HTSbol1,HTSbol5,HTSbol11,HTSbol12,Oktem,HTSbol3,HTSbol6,HTSbol7,Khrebtov,Tkachenko,Nivelle,Kaiser,Frenkel}. Bolometers are often used, e.g.,~for~thermal infrared cameras (see, e.g.,~\cite{HTSbol9,HTSbol4,HTSbol1,HTSbol5,HTSbol11,Oktem,HTSbol6,HTSbol7,Tkachenko,Kaiser}), mm-wave sensing~\cite{HTSbol9,HTSbol4,HTSbol1,HTSbol5,HTSbol11,Oktem,HTSbol7,Tkachenko,Kaiser,HTSbol8,Nivelle,HTSbol12}, space-based~\cite{HTSbol8,Nivelle}, laboratory~far-infrared spectroscopy~\cite{Nivelle}, etc.~\cite{HTSbol9,HTSbol8,HTSbol4,HTSbol1,HTSbol5,HTSbol11,HTSbol12,Oktem,HTSbol3,HTSbol6,HTSbol7,Khrebtov,Tkachenko,Nivelle,Kaiser,Frenkel}. Superconductors are among the best candidate materials for bolometers, due to their extreme sensitivity to $T$ near the superconducting transition, measurable for instance through the sharp variations of the electrical resistance $R$ (resistive transition-edge bolometer---TES). For~resistive TES bolometers, a~key figure for performance is the so-called ``temperature coefficient of resistance'' (TCR), given by~\cite{HTSbol9,HTSbol8,HTSbol4,HTSbol1,HTSbol5,HTSbol11,HTSbol12,Oktem,HTSbol3,HTSbol6,HTSbol7,Khrebtov,Tkachenko,Nivelle,Kaiser,Frenkel,Harry, Irwin,Irwinlibro}:
\begin{equation}
\mbox{TCR}= \left| \frac{1}{R} \frac{\dd R}{\dd T} \right|.
\end{equation}

High bolometric sensitivity requires a large value of TCR. For~instance, structures of vanadium oxides $\mbox{V}_x\mbox{O}_y$,  commonly used in semiconductor-based bolometers, present $\mbox{TCR}\sim  0.025\,\mbox{K}^{-1}$~\cite{Abdel}. Much larger TCR  may be achieved with superconductor materials kept at base temperatures coincident with their normal--superconducting transition, $T_c$. This is the case mainly when  using conventional low-temperature superconductors with $T_c \leq 1$ K (the so-called low-$T_c$ TES bolometers), that achieve TCR $\sim 1000$ K$^{-1}$ or even more~\cite{Harry, Irwin,Irwinlibro}, making them a technology of choice for detecting the most faint radiations, as~the cosmic infrared background or in quantum entanglement and cryptography applications~\cite{Harry, Irwin,Irwinlibro}. Note that for these measurements the very low  temperature required to operate the low-$T_c$ TES is often not seen as a major problem, because~cryogenizing the sensor below a few Kelvin is  required anyway in order to minimize the thermal noise coming from the bolometer itself. However, the~requirement of a highly-stabilized liquid-helium-based cryogenics is a serious difficulty for adoption of low-$T_c$ TES in other~applications. 

After the discovery of high-$T_c$ cuprate superconductors (HTS), various authors have explored their use for resistive bolometers with simpler liquid-nitrogen-based cryogenics (the so-called resistive HTS TES bolometers~\cite{HTSbol9,HTSbol8,HTSbol4,HTSbol1,HTSbol5,HTSbol11,HTSbol12,Oktem,HTSbol3,HTSbol6,HTSbol7,Khrebtov,Tkachenko,Nivelle,Kaiser,Frenkel}). The~compound $\mbox{YBa}_2\mbox{Cu}_3\mbox{O}_{\delta}$ (\YBCO) is the HTS material usually considered for this application, usually with maximum-$T_c$ doping, \ie, stoichiometry $\delta \simeq 6.93$. Such \YBCO\ thin films provide $\mbox{TCR}\sim1.5\,\mbox{K}^{-1}$, low noise at an operational temperature $T_c \sim90$\,K, and~also favorable values for the rest of parameters contributing to good bolometric performance (thermal conductivity, infrared absorbance, response time, etc.)~\cite{HTSbol9,HTSbol8,HTSbol4,HTSbol1,HTSbol5,HTSbol11,HTSbol12,Oktem,HTSbol3,HTSbol6,HTSbol7,Khrebtov, Tkachenko,Nivelle,Kaiser,Frenkel}.

Besides the cryogenics, the other difference with respect to low-$T_c$ TES is that in actual implementations~\cite{HTSbol9,HTSbol8,HTSbol4,HTSbol1,HTSbol5,HTSbol11,HTSbol12,Oktem,HTSbol3,HTSbol6,HTSbol7,Khrebtov, Tkachenko,Nivelle,Kaiser,Frenkel} the resistive HTS TES operate  under current bias and usually in the ohmic regime (instead of non-ohmic resistance 
and the voltage bias employed in low-$T_c$ TES to avoid thermal runaways~\cite{Harry,Irwin,Irwinlibro}).

However, the~HTS TES  until now proposed  still share some of the significant shortcomings of low-$T_c$ TES: First, thermal stability of the cryogenic bath is still challenging (liquid-nitrogen systems are simpler but tend to thermally oscillate more than those based on liquid helium). Secondly, both types of TES have useful TCR only at the superconducting transition, corresponding to operational temperature intervals $\Delta T$ of just $\sim$0.1\,K or less for low-$T_c$ TES, and~$\sim$1\,K for the resistive HTS TES proposed until now~\cite{HTSbol9,HTSbol8,HTSbol4,HTSbol1,HTSbol5,HTSbol11,HTSbol12,Oktem,HTSbol3,HTSbol6,HTSbol7,Khrebtov,Tkachenko,Nivelle,Kaiser,Frenkel,Harry, Irwin,Irwinlibro}.

The HTS TES systems proposed until today are homogeneous in nominal composition and critical temperature~\cite{HTSbol9,HTSbol8,HTSbol4,HTSbol1,HTSbol5,HTSbol11,HTSbol12,Oktem,HTSbol3,HTSbol6,HTSbol7,Khrebtov, Tkachenko,Nivelle,Kaiser,Frenkel}. However, in~the recent years different novel techniques have been developed  to impose regular patterns on HTS thin films, creating custom designs, down to the micro- and the nano-scales~\cite{Villegas,Katzer,Johansen,BendingONE, BendingTWO, BendingTHREE, BendingFOUR,networksONE,networksTWO,Verde}. This allows custom-engineering regular variations of the critical temperature over the film surface. Realization of these regular and controlled patterning has been experimentally achieved using, \eg,  local ferroelectric field-effect~\cite{Villegas}, nanodeposition~\cite{Katzer}, focused ion beam~\cite{Johansen}, etc. In~fact, the nanostructuring of HTS has become the specific subject of recent conferences~\cite{BendingONE, BendingTWO, BendingTHREE, BendingFOUR} and networks~\cite{networksONE,networksTWO} funded by the European~Union.

However, the~use of  nanostructured films for optimizing HTS TES has been considered only very marginally up to now, the~only precedent to our knowledge being Reference~\cite{Oktem} by \mbox{Oktem~et~al.}, who~considered films with random distributions of nonsuperconducting incrustations producing limited increases of $\Delta T$ up to only $\sim$2 K (and also small, and~not always favorable, TCR variations). 

Our aim in the present work is to propose that certain custom nanostructurating and patterning of HTS materials may improve their functionality for resistive HTS TES sensors. In~particular, we~calculate the case of nanostructuring and patterning of the {local carrier doping level $p$ (the number of carriers per CuO$_2$ unit cell) in the prototypical HTS compound YBCO via local variations of oxygen stoichiometry (as realizable, \eg, via local desoxigenation, ion bombardement with different masks, etc.).}  Our main objective will be to obtain an increase of the operational temperature interval, $\Delta T$, in~which {\it i)}~ the TCR is large and {\it ii)}~$R$ is linear with $T$ (\ie, $\dd R/\dd T$ constant with $T$, that is another desirable feature that simplifies both the electronic control of the bolometer and the required stability of the cryogenic setup). Accompanying this $\Delta T$ increase we will also obtain improvements of other bolometric characteristics, such as the saturation power and in some cases the TCR~itself.

We organize our studies of the structured HTS materials in two parts: First we study the simplest case of {carrier} doping nanostructuring, namely the  random nanoscale structuration that appears by just using {oxygen} stoichiometries that do not maximize $T_c$. We present our methods for those randomly structured HTS in Section~\ref{section2}; these consists {of} finite-element computations (and also, to~confirm their validity, analytical estimates using effective-medium approximations~\cite{y}) that we apply to calculate the performance of the material in various example bolometer device implementations. The~results following these methods in random nanoscale structurations are presented in Section~\ref{section4}. These results indicate that this first simple structuration may already improve some of the bolometric parameters with respect to conventional, nonstructured HTS~materials

The rest of the paper considers structurations that include not only the unavoidable random disorder but also the additional imposition of custom regular arrangements of zones with different nominal {doping} levels (patterning), studying different examples aimed to progressively improve the bolometric performance. The~additional methods needed to calculate this added patterning are presented in Section~\ref{section3}, and~the results are discussed in Sections~\ref{constant} to~\ref{discrete} for various custom pattern designs, each of them improving the previous one. The~most optimized pattern design (Section~\ref{discrete}) is a four-step discretized exponential-like dependence of nominal doping with the longitudinal position. This arrangement should be also the easier one to fabricate. With~respect to conventional nonstructured HTS TES materials, it improves by more than one order of magnitude the $\Delta T$ and the saturation power, and~it also doubles the TCR~sensitivity. 

\section{Methods for Structured Nonpatterned Resistive HTS~TES}\label{section2}

Our methodology consists of computing the electrical resistance $R$ versus temperature $T$ of each of the  structured HTS materials considered by us, and~then using such $R(T)$ to calculate the corresponding performance for bolometric~operation. 

For completeness we will, in~fact, consider various example bolometer-device designs, including simple square-shaped sensors such as those in~\cite{HTSbol3, HTSbol6} (that may be a micrometric size as appropriate for building megapixel cameras; we shall also consider two different substrates for completeness) and also larger sensors for millimeter-wavelength sensing using a meander geometry (as those built, always with nonstructured HTS, in~\cite{Tkachenko,Oktem,Nivelle,HTSbol5,HTSbol1,HTSbol7,HTSbol8,HTSbol4,HTSbol9,Kaiser}). While naturally we could not calculate in this work the whole range of possible device designs for a bolometer, our results in these example implementations show that our proposed nano optimizations of the materials should lead to improvements in at least some popular types of resistive HTS TES device~designs.

Also, for~our $R(T)$ calculations we shall use two alternative calculations, so as to be confident about the validity of the results: Finite-element computations first, and~then effective-medium formulae (both paths have been successful in other studies of structured HTS~\cite{Verde,y,Coton2,Coton3,a-benfatto,b-benfatto}, and~we  shall also include some confirming example comparisons with real data). Let us provide the details of all such~procedures in the following sections. 

\subsection{Main Operational Parameters for Resistive HTS TES~Devices}\label{operational}

We consider in this work, a HTS TES of resistive type, \ie, the~temperature increase is sensed through the measurement of the electrical resistance, as~in~\cite{HTSbol9,HTSbol8,HTSbol4,HTSbol1,HTSbol5,HTSbol11,HTSbol12, Oktem,HTSbol3,HTSbol6,HTSbol7,Khrebtov,Tkachenko,Nivelle,Kaiser,Frenkel}. Contrarily to the most common case of low-$T_c$ TES, the~measurement is in current-bias ($I$-bias) configuration in all experimentally implemented resistive HTS TES published to our knowledge (the experimental difficulties for voltage-bias ($V$-bias) sensing in HTS TES were explained, \eg, by~Khrebtov~et~al.~\cite{Khrebtov}). Also the $I$ value usually employed~\mbox{\cite{HTSbol9,HTSbol6,Tkachenko,Kaiser}} is sufficiently small as to correspond to the ohmic regime ($R=V/I$ constant with $I$) in all the operational $T$-range (see also below; this is also in contrast to low-$T_c$ TES). As~already mentioned in the Introduction, for~ohmic resistive TES a main parameter of merit is the TCR, that may be also expressed as
\begin{equation}\label{TCR}
 \mbox{TCR}= \frac{R(T^+) - R(T^-)}{\Delta T \ R(T^-)},
\end{equation}
where $T^-$ is the base operation temperature (the one in absence of radiation), $T^+$ is the maximum temperature up to {which} the ohmic $R(T)$ maintains the strong and constant slope with temperature,~and
\begin{equation}\label{deltaT}
\Delta T =T^+-T^-
\end{equation}
will be henceforth called the operational temperature~interval.

The other important parameter is $P^{\rm max}$, the~maximum power measurable without saturation. In~a $I$-bias resistive TES, it is possible to obtain $P^{\rm max}$ at good aproximation~\cite{Irwin,Irwinlibro} by just applying the heat flow equilibrium condition at saturation:
\begin{equation}\label{Pmax}
P^{\rm max} + I^2 R(T^+) =  G \Delta T ,
\end{equation}
where $I^2 R(T^+)$ is the heat rate due to the Joule effect, $G \Delta T$ is the power dissipated towards the cryobath, and~$G$ is the thermal conductance between the film and the~bath.

We shall consider in this work three example resistive HTS TES device designs, to~probe the effects of our proposed material optimizations in them. In~particular, we consider two cases of microsensor bolometers, plus one case adapted to millimeter-wavelength sensing that we specify in the following section.

\subsubsection{Microsensor Device~Design} \label{microsensor}

The first resistive HTS TES device design consists of depositing a thin layer of YBCO HTS material over a substrate, the~area of the HTS and the substrate being micrometric. In~particular, we consider the convenient area (6 $\upmu$m)$^2$, that makes possible building a 1 megapixel array of sensors in $\sim$1 cm$^2$. Each substrate is in direct contact with a cryogenic liquid-nitrogen bath and we shall consider two possible substrate compositions: SrTiO$_3$ (STO, most appropriate to grow HTS films) and a CMOS-type substrate of interest for technological integration (note that HTS TES over CMOS substrates, in~particular silicon/Yttria-stabilized zirconia (YSZ)/zirconia, have been already fabricated~\cite{Kaiser}, using nonstructured YBCO). The~experimental value of $G$ for both types of substrate can be obtained from~\cite{kappa,Kaiser}. We~consider YBCO thickness 100 nm and substrate thickness 1 mm. Also, we consider a bias current of $I=6~\upmu$A, that corresponds to a current density $j=10^3$ A/cm$^2$, a~value used in~\cite{Nivelle,Khrebtov} and that corresponds to the ohmic range in all the $T$-range of operation~\cite{Puica,Nivelle,Khrebtov}.

\subsubsection{Millimeter-Wave Sensor Device~Design} \label{milisensor}

The second device design we shall consider corresponds to the one most employed by experimentalists having produced resistive HTS TES~\cite{HTSbol9,HTSbol4,HTSbol1,HTSbol5,HTSbol11,Oktem,HTSbol7,Tkachenko,Kaiser,HTSbol8,Nivelle}. It corresponds to a larger design using, as substrate, a suspended membrane of millimetric surface and CMOS-type composition; on top of such thin (micrometric thickness) membrane substrate a single meander of YBCO material is deposited. The~larger area precludes building small megapixel sensors, but~this is not important, \eg, for~sensing {millimeter} wavelengths (that could not be constrained in smaller pixel sizes anyway, and~that are among the main applications of bolometers). The~meander geometry allows instead the ability to optimize the so-called ''static voltage responsivity``, $S_{\rm V}$, an~important parameter defined by:
\begin{equation}
S_{\rm V}=\frac{\varepsilon L_0}{I(1-L_0)},
\end{equation}
where $\varepsilon$ is the absorbance of the sample, $I$ is the bias current, and~\begin{equation} 
L_0=\frac{\rm TCR \; \textit{I}^2 \; \textit{R}}{G}
\label{L0}
\end{equation}
is the so-called loop gain coefficient, which is a relative measure of the positive electrothermal feedback of the device~\cite{Khrebtov} and $G$ is again the thermal conductance towards the bath (whose experimental value may be found, \eg, in~\cite{Nivelle}); we consider (3 mm)$^2$ membranes for that evaluation. For~a stable operation, the~loop gain coefficient $L_0$ should be smaller than $1$, the~value $L_0=0.3$ being usually taken as optimal. Therefore, the~maximum static responsivity is obtained by tuning the geometry of the meander so to tune the $I^{2} R$ contribution to $L_0$ in Equation~(\ref{L0}) (Let us also note here that for $I$-bias the $L_0$ may remain constant, and~$L_0=0.3$, only if $R$ is linear with $T$, so that maximizing $\Delta T$ is also interesting in this~respect).

In our calculations, we will use for the meander section the same size 6$\upmu$m x 100 nm as previously for $\upmu$m-sensors. We then choose the meander length for each sensing material so that always $L_0\sim0.3$. We also consider the same bias current and current density, I = 6 $\upmu$m and $j=10^3$ A/cm$^2$, than~for $\upmu$m-sensors (again corresponding to the ohmic range~\cite{Puica,Nivelle,Khrebtov} and comparable to values used in experimental meander resistive HTS TES~\cite{Nivelle,Khrebtov}). These choices not only are realistic but also they will allow us to use the same computer calculations of $R(T)$ for both $\upmu$m- and mm-device designs (as only a geometric correction prefactor is needed to change $R$ from one design to the other) and the same pair of bias current and current density, $I=6~\upmu$A and $j=10^3$ A/cm$^2$ .

\subsection{$R(T)$ in the Normal State of Nonstructured HTS}\label{Background}

In HTS materials, the ohmic resistance versus temperature, $R(T)$, markedly varies with the doping level $p$ (number of carriers per CuO$_2$ unit cell that for instance in YBCO may be changed through the oxygen content). This is true both for the value of the critical temperature $T_c (p)$ below which, the superconductivity transition occurs, and~for the $R(T)$ magnitude and $T$-dependence in the normal state $T>T_c(p)$. The~$R(T)$-versus-$p$ phase diagram has been extensively studied in many works such as the review~\cite{PNAS} (see also, \eg,~\cite{Nature3,Tafti}). Here, let us recall that the superconducting critical temperature is maximum at $p \sim 0.155$ (separating the so-called underdoped $p<0.155$ and overdoped $p>0.155$ compositions). Above~$T_c(p)$, the~material presents a normal-state background electrical resistivity, $\rho_{\rm b}(T,p)$, that is linear on $T$ above a certain so-called pseudogap temperature $T ^*$, and~is pseudoparabolic semiconducting-like~\cite{PNAS} for $T_c<T<T^*$. In~YBCO, it is $T^*(K) \approx 270-3000(p-0.1)$~\cite{PNAS}, so that for $p \gtrsim 0.16$, it is $T^*<T_c$ and the semiconducting-like region disappears. Instead of these rapid crude approximations, we will use in our analysis, all through the present work, the~detailed quantitative results for $T_c(p)$, $T^*(p)$, and $\rho_{\rm b}(T,p)$ given in reference~\cite{PNAS} for~YBCO.

Near $T_c(p)$, obviously $R(T)$ undergoes the superconducting transition towards $R(T)=0$. This~transition is not fully sharp, instead, a~sizable rounding of $R(T)$ occurs in the vicinity of $T_c(p)$. This rounding is known to have two contributions: critical fluctuations and doping inhomogeneities, that we describe in the following~subsections.

\subsection{Rounding of $R(T)$ Near the Superconducting Transition Due to Critical Phenomena}\label{rounding}

The critical fluctuations around the transition play an important role in HTS and have been studied in detail, \eg, in~\cite{Carballeira,Ramallo,Mosqueira,Coton,Coton2,Solov,a,b,c}. The~effects of critical fluctuations in the resistance curves are commonly summarized via the so-called paraconductivity, $\Delta \sigma$, defined as the additional contribution to the electrical conductivity due to fluctuations: In particular, the~total conductivity $\sigma(T)$ near the transition~becomes
\begin{equation}\label{rhoeq}
\sigma= \Delta \sigma+1/\rho_{\rm b}, 
\end{equation}
where $\rho_b$ is the normal-state background resistivity (see previous subsection). Because~$\Delta \sigma$ follows critical-divergence laws near the transition, its effect far from $T_c$ (for $T \gtrsim 1.7 T_c$) is totally negligible. Closer to $T_c$, however, $\Delta \sigma$ becomes progressively important and two $T$ ranges may be distinguished. For~$1.01T_c \lesssim T \lesssim 1.7T_c$, \ie, the~so-called Gaussian fluctuations region, $\Delta \sigma$ is well described by the Lawrence--Doniach paraconductivity equation for layered superconductors:~\cite{Carballeira, Ramallo, LD,a,b,c}
\begin{equation} \label{GGL}
\Delta \sigma = \frac{e^2}{16 \hbar d} \left\{  \frac{1}{t}\left(1+\frac{B_{\rm LD}}{t} \right)^{-1/2}-\frac{2}{c}+\frac{t+B_{\rm LD}/2}{c^2}  \right\},
\end{equation}
where $e$ is the electron charge, $\hbar$ is the reduced Planck constant,  $t=\ln (T/T_c)$ is the reduced temperature, $B_{\rm LD}=(2 \xi_c(0)/d)^2$ is the Lawrence--Doniach~\cite{LD} layering parameter, $\xi_c(0)$ is the Ginzburg--Landau coherence length amplitude in the out-of-plane direction ($\simeq$1.1~\AA~ in YBCO~\cite{Ramallo, Carballeira}), $d$~is the superconducting layer periodicity length ($\simeq$5.85~ \AA~  in YBCO~\cite{Ramallo, Carballeira}) and  $c$ is a high-temperature cutoff constant $\simeq$0.7~\cite{Carballeira,b,d,e}.

Closer to $T_c$, for~$T_{BKT}\lesssim T \lesssim 1.01T_c$ we find the strong phase fluctuation regime, dominated by the Berezinskii--Kosterkitz--Thouless (BKT) transition temperature $T_{BKT}$ ($\sim T_c -  2K$ in YBCO)~\cite{Ying,c}. In~this regime, the paraconductivity can be obtained using the equation:~\cite{Coton, Coton3,Halperin,a-benfatto,b-benfatto}
\begin{equation} \label{BKT}
\Delta \sigma=A_{\rm BKT} \ {\rm exp} \left[ 4 \sqrt{\frac{ T_c-T_{BKT}}{T-T_{BKT}}} \ \right],
\end{equation}
where $A_{\rm BKT}$ is a constant, obtainable by requiring continuity of Equations~(\ref{GGL}) and ~(\ref{BKT}) at the intersection of the Gaussian and BKT regimes, \ie, at~$T=1.01 T_c$.

\subsection{$R(T)$ Transition Rounding Due to Intrinsic Structuration of the {Carrier} Doping Level; Nominal vs. Local~Doping}\label{inh}
As it has been explicitly demonstrated in various relatively recent experimental and theoretical works~\cite{Mosqueira, Mosqueira2, Mihailovic, Coton, Coton2}, an~additional (and crucial for some doping levels) ingredient to understand the phenomenology of the resistive transition in HTS is to take into account the random $T_c$-inhomogeneities associated with the intrinsic disorder of the doping level. This intrinsic structuration is due to the fact that HTS compounds are non-stoichiometric, and~therefore each dopant ion has at its disposal various lattice positions to occupy. {For concreteness, we focus our present article in the case of the YBa$_2$Cu$_3$O$_\delta$ superconductor with oxygen as a dopant ion.} Experimental measurement indicates that a typical size of each local inhomogeneity is about ($30$ nm)$^2$ for HTS~\cite{Mihailovic, Mosqueira2,Coton2}. This produces, therefore, a~certain randomness in the doping at the local scale, unavoidably present in even the more carefully grown HTS samples. A~relatively easy geometrical calculation~\cite{Mosqueira2, Mihailovic, Coton2} reveals that this intrinsic structuration shall produce a Gaussian distribution of local dopant levels, as~\begin{equation}\label{omega}
\omega({p},{\overline{p}})=\frac{2\sqrt{\ln 2}}{\sqrt{\pi}\Delta {p}} \exp \left[ - \left( \frac{{p}-{\overline{p}}}{\Delta  p/(2\sqrt{\ln 2})} \right)^2 \right], 
\end{equation}
where $\omega(p,\overline{p})$ is the fraction distribution of local doping levels, $p$, for~a film with average doping level $\overline{p}$ (henceforth called nominal doping), and~$\Delta p$ is the FWHM of the Gaussian distribution. This~$\Delta p$ may be obtained, in~turn, on~the grounds of coarse-graining averages (see, \eg, Equation~(6) of References~\cite{Coton2,Mosqueira2}) and for YBCO it is $\Delta p \sim 0.006$ (with a small dependence on $\overline{p}$ that may be considered in excellent approximation linear $\Delta p = 0.0032 + 0.0189 \overline{p}$)~\cite{Coton2,Mosqueira2}.

Due to the $T_c(p)$ dependence in HTS, the~above distribution of local $p$ values leads, in~turn, to~a corresponding distribution of local critical temperatures around the nominal value $\overline{T}_c=T_c(\overline{p})$. The~corresponding full width at half maximum (FWHM) for such intrinsic $T_c$ structuration has been also considered, \eg, in~\cite{Mosqueira2,Coton2}. Not surprisingly, it becomes quite negligible ($\sim$1 K in YBCO) for the nominal dopings $\overline{p}\simeq 0.155$ that maximize $T_c$ (and that has been used up to now for HTS TES; $\overline{p}\simeq 0.155$ corresponds to YBa$_2$Cu$_3$O$_{6.93}$ stoichiometry at which $T_c(p)$ is maximum and $dT_c/dp \sim 0$). However, for~other dopings the situation may become very different and the $T_c$ distribution can reach FWHM values as large as, \eg, $\sim$5 K for $\overline{p} = 0.13$, significantly influencing the $R(T)$ roundings~\cite{Mihailovic, Mosqueira2,Coton2}.

\subsection{Obtainment of the $R(T)$ Curve of Nonpatterned HTS TES Using Finite-Element Computations}\label{FEl}

To calculate the resistance transition curves, $R(T)$, of~the $T_c$-structured HTS material, we have used software (TOSERIS, available by request to authors) that numerically solves the electrical mesh-current matrix equations of a film modeled as a $200 \times 200$ square lattice of monodomains, where each domain $i$ may have its own doping $p_i$, and~thus its own $T_{ci}$ and resistivity curve. We have used Equations~(\ref{rhoeq})--(\ref{BKT}) for the $\rho_i(T,p)$ functionality of each monodomain $i$. The~model also includes an I-bias power source and a voltmeter connected with zero-resistance contacts to opposite edges of the sample (see, \eg, scheme in Figure~\ref{figA}) and the $R(T)$ of the film will be obtained as the external $V/I$. Calculating the circuit requires to numerically invert, for~each temperature, the~sparse matrix with dimensions $40,001\times40,001$ that defines the mesh-current equations. This is a parallelizable computation for which we employed a 31 Tflops supercomputer (LBTS-$\varepsilon$psilon, that comprises about 12,000 floating-point units and is described in~\cite{lbts}). It was 100\% allocated to run our software during several~weeks.

 We have performed our calculations with numerical values representative of the HTS compound YBCO and therefore for the area of a finite-element monodomain $i$ we used (30 nm)$^2$, that is expected to correspond to the size of a doping $T_c$ inhomogeneity in YBCO.~\cite{Mihailovic,Mosqueira2,Coton2} Therefore, the~surface of the simulated HTS film is going to be (6 $\upmu$m)$^2$, in~agreement with the microsensor HTS TES device implementation of Section~\ref{microsensor}.
 
In the case of the nonpatterned HTS considered in this section, the~only spatial variation of doping and $T_c$ is the unavoidable intrinsic dopant ion structuration and, thus, we assign the local $p_i$ and $T_{ci}$ value to each of our $200 \times 200$ monodomains $i$ as follows: We first build a set of 40,000 values of dopings following the Gaussian distribution given by Equation~(\ref{omega}). We then assign each of those $p$-values to each node $i$ randomly. Finally, those $p_i$ are transformed to $T_{ci}$ values (and corresponding $\rho_i(T)$ functions) following the quantitative results of~\cite{PNAS}. A~scheme of an example of the resulting spatial distribution is provided in Figure~\ref{figA} (note the random nanostructuration in the zoomed area).

\subsection{Analytical Estimates Using an Effective-Medium~Approximation}\label{EMap}
To additionally probe the consistency of our computations, we will use, as~a useful test, semi-analytical results that we calculate using the so-called effective-medium equations (EM approximation). The~EM approximation was first introduced by Bruggeman~\cite{x} for general random inhomogeneous materials, and~then adapted, \eg, by~Maza and coworkers~\cite{y} for HTS with Gaussian random $T_c$ distributions. As~shown in those early works, the~EM approximation is a coarse-averaging model that may be considered accurate for temperatures not too close to the $R \rightarrow 0$ point (at which percolation effects may be expected to invalidate the approximation). In~the case of our 2D media, the~EM equations may be summarized as the following implicit condition for the conductivity $\sigma$ of each region with random doping inhomogeneities~\cite{y,a-benfatto,b-benfatto,Coton3}:
\begin{equation} \label{inhom}
\int_{0}^{\infty} \frac{\sigma_{p}-\sigma}{\sigma_{p}+ \sigma} \ \omega({p},{\overline{p}}){\rm d} p=0.
\end{equation}
Here, $p$, $\overline{p}$ and $\omega(p,\overline{p})$ retain the same meaning as in Equation~(\ref{omega}), and~$\sigma_p$ is the electrical conductivity corresponding to doping level $p$. The~above equation has to be numerically solved to obtain $\sigma$; however, the~computational weight is much lower than the finite-element computation method (seconds versus hours or even days in our parallel computer~\cite{lbts}).

\section{Additional Methods for Structured and Patterned Resistive HTS~TES}\label{section3}

We now describe the additional methods needed to obtain the $R(T)$ curve of HTS films in which, additional to the random nanostructuration considered in previous section, also a regular pattern of nominal doping levels is imposed, with~the aim to obtain designs that optimize the bolometric functionality. In~these films, a~regular spatial variation of the nominal doping level $\overline{p}$ is created by the samples' grower by using any of the different methods for micro- and nanostructuration developed in the recent years by experimentalists in HTS films (see, \eg,~\cite{Villegas,Katzer,Johansen,BendingONE, BendingTWO, BendingTHREE, BendingFOUR,networksONE,networksTWO,Verde}; for instance, for~YBCO, this is possible by local deoxygenation using cover masks, ion bombardment, etc.). In~particular, all the specific example patterns considered in this work will be expressible as functions $\overline{p}(x)$, where $x$ is the coordinate in the direction parallel to the external bias current (see scheme in Figure~\ref{figA}) and thus it will be useful to introduce the corresponding function $\lambda(\overline{p})$, or~relative length weight of each nominal $\overline{p}$ value in the film, defined as
\begin{equation} \label{lambda}
\lambda(\overline{p})=\frac{1}{L}\frac{dx}{d\overline{p}},
\end{equation}
where $L$ is the total length of the film in the $x$-direction. Crucial for our studies, one has still to add to these nominal $\overline{p}(x)$ variations the unavoidable nanometric-scale randomness of the doping level (considered in the previous sections), \ie,
\begin{equation}
 p(x,y)=\overline{p}(x)+ p_{\rm random} (x,y),                                                                                                                                                                                                                                                                                                                                                                                                                                                                                                                                                                                                                                                                                                                                                                                                                                                                                                                                                                                                                                                                                                                                                                                                                                                                                                                                                                                                                                                                                                                                                                                                                                                                                                                                                                                                                                                                                                                                                                                                                                                                                                                                                          
\end{equation}
with $p_{\rm random}(x,y)$ consistent with Equation~(\ref{omega}) evaluated using the local $\overline{p}(x)$.

\subsection{Obtainment of the $R(T)$ Curve of Patterned HTS Using Finite-Element Computations}\label{FEs}

To obtain the $R(T)$ curves of patterned resistive HTS TES, we use the finite-element software TOSERIS already described in Section~\ref{FEl}. We again use a $200\times200$ simulation mesh and now we assign to each of those finite elements a local doping as follows: First, we associate to each element $i$ a nominal doping $\overline{p}_i$ corresponding to the pattern to be simulated. Then we randomly calculate the local doping $p_i$ following the Gaussian distribution given by Equation~(\ref{omega}), evaluated with the nominal doping $\overline{p}_i$ of each node. Finally, to~each node we assign the $T_{ci}$ and $\rho_i(T)$ corresponding to their local $p_i$ as per the quantitative results of~\cite{PNAS} for the HTS material YBCO (see Sections~\ref{Background} and~\ref{rounding}).

We also tested that the sets of nodes sharing the same $\overline{p}_i$ value follow the Gaussian distribution, and~each $R(T)$ simulation was repeated for several so-generated samples to verify their reproducibility. These checks indicate that our choice  of a $200\times200$ node  mesh provides enough statistical size. If~we attribute to each node the size $(30~$nm$)^2$ corresponding to each $T_c$-monodomain in YBCO (see~Section~\ref{inh} and~\cite{Coton2,Mosqueira2}), the~whole $200\times200$ sample corresponds to $(6~\upmu$m$)^2$, that is realistic for a microbolometric~pixel.  

Unless stated otherwise, we will again use in our calculations  the numerical values in Sections~\ref{Background} to~\ref{FEl} for the common material characteristics, such as, \eg, a~film thickness of 100 nm or values for the critical-fluctuation parameters as per Section~\ref{rounding}.

\subsection{Analytical Estimates Using an Extended-EM~Approximation}\label{EMpat}
Besides performing finite-element computations, we will test our results  against estimates based on the EM approach. For~that purpose, we must suitably extend this approximation to account for the 1D gradient of nominal dopings corresponding to each example pattern to be considered in this work. For~that, we consider the film as an association in a series of domains, each one with its own resistance and nominal doping $\overline{p}$, so that:
\begin{equation}\label{Ra}
R(T)=\int_{x=0}^{x=L}\frac{{\rm \textrm{d} \overline{p}}}{\sigma({ \overline{p}(x)},T)S},
\end{equation}
where $L$ is again the total length of the superconductor, $S$ is its transversal surface, and~$\sigma(\overline{p}(x),T)$ is the ohmic conductivity obtained using the monodomain-EM approach, \ie, using Equation~(\ref{inhom}) for each doping $\overline{p}$(x). Equation~(\ref{Ra}) can be also written as the following integration over nominal doping, with~the help of the $\lambda ( \overline{p})$ function defined in Equation~(\ref{lambda}):
\begin{equation}\label{Rlam}
R(T)=L\int_{\overline{p}_0}^{\overline{p}_L} \lambda (\overline{p}) \frac{{\rm d \textit{x}}}{\sigma({ \overline{p}(x)},T)S},
\end{equation}

For a discrete distribution (stepwise function $\overline{p}(x)$), the~above equation becomes instead a~summation:
\begin{equation}
 R(T)=\sum_{i=1}^{N} \frac{L_i}{\sigma(\overline{p}_i,T)S},
 \label{Rde1}
 \end{equation}
 where $N$ is the number of discrete domains, each with its own nominal doping $\overline{p}_i$ and length $L_i$. 
 
 Note that Equations~(\ref{Ra}) to (\ref{Rde1}) do not explicitly take into account the transverse currents when associating the different domains $\overline{p}_i$ and $L_i$. Non-longitudinal transport inside each domain is built-in by using the EM approach for each $\sigma(\overline{p}_i,T)$. However, this approximation may be expected to fail when it has to describe percolations (because both the sum in a series of domains and the EM approximation do not take them into account). Therefore it could be expected to overestimate the value of $R(T)$ in the very close proximity to the fully superconducting $R(T) \rightarrow 0^{+}$ state.

\section{Results for Structured Nonpatterned HTS~Materials}\label{section4}

In the reminder of this article, we describe the results of applying our methods to different structured HTS materials. We consider first the case of nonpatterned HTS materials, \ie, those with a uniform nominal doping $\overline{p}$. As~already mentioned, the~doping level $\overline{p}\simeq 0.155$ (corresponding to the stoichiometry YBa$_2$Cu$_3$O$_{6.93}$) is the one that has been used up to now to experimentally produce HTS TES, and, due to the saturation of $T_c$ near such doping level, it corresponds to a HTS film without a $T_{c}$-nanostructure. However, the~cases with uniform $\overline{p}< 0.155$ correspond to films with random $T_{c}$-nanostructuration.

\subsection{Results for the $R(T)$ Profile and Operational Parameters for Resistive TES Case}\label{resultA}

\begin{figure}[!ht]
\centering
\includegraphics[]{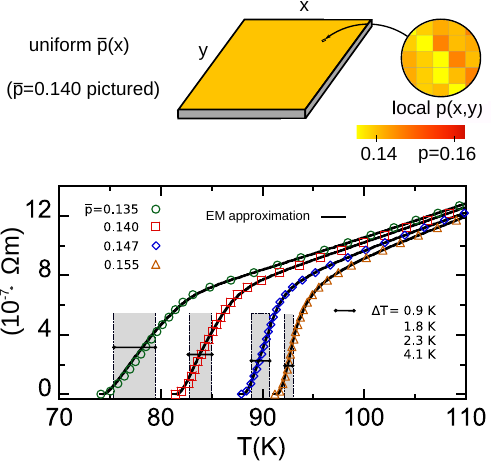}
\caption{Electrical resistivity $\rho$ versus temperature $T$ obtained for $\mbox{YBa}_2\mbox{Cu}_3\mbox{O}_{\delta}$ (\YBCO) films with a single, uniform value for the nominal doping level $\overline{p}$, including the case with negligible $T_c$ nanostructuration (or~maximum-$T_c$ doping, $\overline{p}=0.155$, in~which $T_c$ saturates near its maximum value and the $T_c$ disorder is negligible) and various cases in which the $\overline{p}$ value corresponds to significant $T_c$ nanostructuration (see Section~\ref{section4} for details). The~data points correspond to the finite-element computations, and~the continuous lines to the analytical effective-medium (EM) equations. The~shaded gray region signals the operational temperature range $\Delta T$ (in which $\rho$ is strongly dependent and linear in $T$). In~the upper drawing we illustrate the simulated sample and setup, also including  a zoom at smaller length scales, imaging the spatial variation of the local doping level $p(x,y)$ (each $p(x,y)$ monodomain has typical size (30~nm)$^2$ and the distribution is Gaussian around the average $\overline{p}$(x), see main text for details).}
\label{figA}
\end{figure}

In Figure~\ref{figA} we show the $R(T)$ resulting from our finite-element computations for the case of YBCO films with nominal dopings $\overline{p}=0.135,0.140,0.147$ and $0.155$, represented as circles, squares, diamonds, and triangles, respectively. As~previously mentioned, to~these nominal dopings,  a random intrinsic structuration has to be added (following a Gaussian distribution as per Equation~(\ref{omega})) in order to obtain the local $p(x,y)$ values; this is illustrated by the zoom in the pictured drawn in Figure~\ref{figA}.

As it is clearly shown in that figure, the~reduction of $\overline{p}$ induces not only a shift of the transitions towards somewhat lower temperatures (as expected from the $T_c$-vs-$\overline{p}$ phase diagram of HTS), but~also a widening of the $T$-width of the transition region. The~gray areas in Figure~\ref{figA} are the $T$-range, $\Delta T$, in~which the $R(T)$ transition occurs and $R$-vs-$T$ is linear, \ie, the~operational interval $\Delta T$ of Equation~(\ref{deltaT}). This $\Delta T$ increases as $\overline{p}$ decreases, as~may be noticed both in Figure~\ref{figA} and also in Table~\ref{table1}. In~this table we summarize the bolometric operational parameters obtained by using such $R(T)$ results and the methods in Section~\ref{section2} for three example resistive HTS TES devices designs (described in Sections~\ref{microsensor} and~\ref{milisensor}).

It is clear that the improvement that the $T_c$-nanostructure may provide for such bolometric characteristics: Notably, the~increase of $\Delta T$ (about five times higher for the more nanostructured case than in the nonstructured one) is translated in an enhancement of $P^{\rm max}$ of the bolometer device.  This~means the ability to receive a higher amount of radiation without saturating the sensing material. Note that a larger $\Delta T$ also implies a less demanding cryogenic setup in terms of required~stability. 

Therefore, our studies suggest that this first simple $T_c$ nanostructuration already improves some of the bolometric operational parameters with respect to the conventional, nonstructured HTS materials proposed until~now.

\subsection{Verification Using the Analytical EM Approximation and Against Existing~Measurements} \label{EM}

\begin{figure}[!ht]
\centering
\includegraphics[]{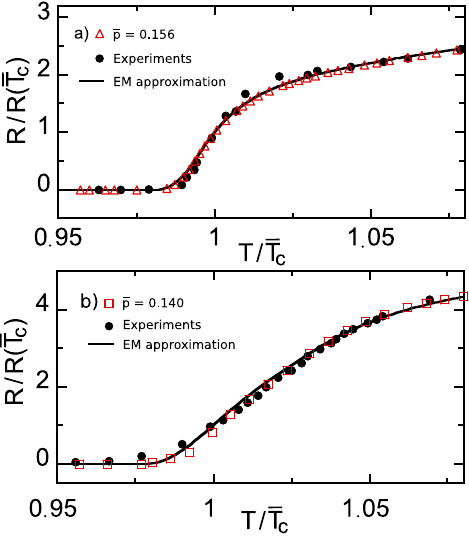} 
\caption{{Comparison between} the electrical resistance vs. temperature curves resulting from our methodology (open symbols for the finite-element computations and solid line for the EM analytical approximation) and the measurements of reference~\cite{Solov} (solid symbols), for~YBCO films with uniform nominal oxygen stoichiometry corresponding to: in panel a) to $\overline{p}=0.156$, \ie, very near the maximum-$T_c$ doping level; in panel b) to $\overline{p}=0.140$, \ie, a~$T_c$-nanostructured nonpatterned HTS. Note~that the normalization of the axes significantly zooms the transition region with respect~to~Figure~\ref{figA}.}
\label{Figure1}
\end{figure}

Figure~\ref{figA} also shows (as continuous lines) the results obtained by applying the EM approach, \ie, Equation~(\ref{inhom}), to~the same parameter values and doping levels as used in the previous subsection. It can be seen in that figure that the coincidence between the finite-element computation and the EM approximation is excellent, even in the $R \rightarrow 0^{+}$ tails (a log--log zoom of such tails evidences moderate deviations in relative values, negligible in the absolute scale of Figure~\ref{figA}, as~coherent with the expectation that the EM approximation is less accurate when percolative current paths appear~\cite{x,y}; see also Figure~\ref{Figure1} for better evidence of these $R \sim 0^+$ deviations. This comparison gives, then, a~first argument supporting the validity, at~least concerning the main features, of~our calculation methods for the doping structuring~effects. 

A second argument supporting such validity is the comparison with actual measurements for $R(T)$. In~the case of structured nonpatterned HTS films considered in this section, measurements valid for such a comparison do exist. In~particular, we have plotted in Figure~\ref{Figure1} the data measured in reference~\cite{Solov} in high-quality YBCO (YBa$_2$Cu$_3$O$_{\delta}$) films comprised by a single zone of nominal oxygen stoichiometries $O_{6.78}$ and $O_{6.85}$ (\ie, $\overline{p}\simeq 0.140$ and $\overline{p}\simeq 0.156$ respectively; to get the relations between oxygen ratio and doping we have interpolated the experimental data described in reference~\cite{Tallon}). We can see in that figure the good accuracy of the theoretical (EM approach) and computational (finite-element) methods used in this article to reproduce the experimental resistance curves of nonpatterned YBCO films in the studied doping~range.

\section{Results for Structured HTS Materials Patterned with a Linear $\overline{p}(x)$ Variation}\label{constant}

Let us now study different instances of doping-patterned HTS, seeking to progressively identify pattern designs optimizing the performance as bolometric sensor materials. We start by considering in this section, a simple linear variation of $\overline{p}$ along the longitudinal direction (the direction of the overall externally-applied electrical current, see the scheme in Figure~\ref{fig3}):
\begin{equation}\label{linear}
\overline{p}(x)=\overline{p}_{0} + \left(\frac{\overline{p}_{L}-\overline{p}_{0}}{L}\right)x,
\end{equation}
where $\overline{p}_{0}$ and $\overline{p}_{L}$ are the $\overline{p}$-values at the opposite ends of the film $x=0$ and $x=L$. In~terms of the $\lambda(\overline{p})$ function of Equation~(\ref{lambda}), this linear-in-$x$ $\overline{p}$-pattern simply becomes the constant value
\begin{equation}
\lambda(\overline{p})=\frac{1}{\overline{p}_{L}-\overline{p}_{0}}.
\label{lambdacons}
\end{equation}
For our computations we chose the rather typical values $\overline{p}_{0}=0.135$ and $\overline{p}_{L}=0.161$.

\subsection{Results for the R(T) Profile and Operational Parameters for Resistive TES Use}

\begin{figure}[!ht]
\centering
\includegraphics[width=8 cm]{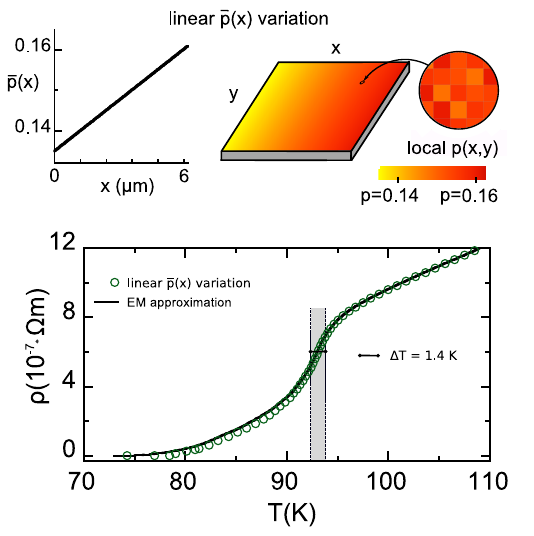}
\caption{In the upper row, we illustrate a YBCO film patterned following the linear variation of the nominal doping $\overline{p}(x)$ studied by us in Section~\ref{constant} (see Equations~(\ref{linear}) and~(\ref{lambdacons})). We also illustrate doping across the film as a 2D color map, taking into account that the local doping level $p(x,y)$ (zoom in the picture) results from accumulating the lineal $\overline{p}(x)$ and the random Gaussian disorder at smaller length scales of about (30 nm)$^2$ (see main text for details). In~the lower row, we plot the resistivity vs. temperature $\rho(T)$ that we obtain for such film (data points for finite-element computations, continuous line for the analytical EM approximation, see Equation~(\ref{eq20})). Note that the transition widens considerably with respect to Figure~\ref{figA}, but~the operational $\Delta T$ range (shaded gray region) is small due to the nonlinearity of $\rho$ with $T$ in most of the~transition. }
\label{fig3}
\end{figure}

The results of our numerical finite-element evaluation for this $\overline{p}$-pattern are displayed in Figure~\ref{fig3} (see also Table~\ref{table1} for a comparative summary). As~evidenced in the Figure~\ref{fig3}, this type of structuring of the HTS film significantly broadens the $R(T)$ transition (compare, \eg, with~Figure~\ref{figA} that represents, in~the same $T$-scale, the~results for HTS with comparable, but~uniform, $\overline{p}$-values). However, this~structuring does not lead to a linear dependence of $R$ vs. $T$ in that transition region. This~may pose a difficulty in resistive TES applications, that ideally require a $R(T)$ variation both large (\ie, large TCR) and linear (\ie, constant $\textrm{d}R/\textrm{d}T$). The~range $\Delta T$ in which both conditions are met is merely about $1.4~\textrm{K}$ for this type of $\overline{p}$-pattern, already suggesting that further structuring optimizations would be desirable (see next section). In~Table~\ref{table1}, we summarize the operational parameters obtained for the linear $\overline{p}(x)$ resistive bolometer. We can conclude that they are of the same order or worse than the parameters obtained for the typical already existing resistive HTS TES (the case $\overline{p} \simeq 0.155$). This~even includes the TCR, that is lower due to the increase of the operational temperature $T^-$ and then of $R(T^-)$. The~maximum energy and power may be somewhat higher due to the small increase of the width of the linear~regime.

We can conclude that this first patterning does not effectively optimize the operational parameters of the resistive HTS TES, mainly because it broadens the $R(T)$ transition but does not achieve $R(T)$ linearity in~it.

\subsection{Verification Using the Extended-EM Analytical~Approximation}\label{EM2}
To check the validity of our results for the linear $\overline{p}(x)$ variation, we also have used the formulae described in Section~\ref{EMpat}. For~that, our first step has been to combine Equation~(\ref{Rlam}) with the $\lambda (\overline{p})$ formula for this type of pattern (Equation~(\ref{lambdacons})). This leads us to the new equation:
\begin{equation}
R(T)=   \frac{L}{S(\overline{p}_L-\overline{p}_0)}  \int_{\overline{p}_0}^{\overline{p}_L}  \frac{{ d\overline{p}} }{\sigma({ \overline{p}},T)} \ \ \ {\rm (linear} \  \overline{p}(x) \ \rm{pattern)},
\label{eq20}
 \end{equation}
where $\sigma(\overline{p},T)$ results from Equation~(\ref{inhom}). The~result of this analytical estimate is displayed in Figure~\ref{fig3} as a continuous line. As~in the case described for constant nominal doping, this estimate achieves good agreement with the finite-element computation, confirming the basic accuracy of our~results.

\section{Results for Structured HTS Materials Patterned with a Continuous Exponential-Like Doping~Variation}\label{exponential}

Seeking to find a $\overline{p}(x)$ profile producing a $R(T)$ transition that improves the bolometric operational characteristics, we have explored numerous $\overline{p}(x)$ options beyond the simple linear function discussed above. In~the present section we present the results that we obtained with the continuous $\overline{p}(x)$ functionality that led us to better bolometric performance (and a step-like, noncontinuous variation will be later discussed, in~Section~\ref{discrete}). This continuous $\overline{p}(x)$ profile is more intuitively described by means of the length weight function $\lambda(\overline{p})$. In~particular, we consider $\overline{p}$-profiles leading to the following exponential $\lambda(\overline{p})$ function:
\begin{equation}
\lambda(\overline{p})=A\exp\left(\frac{\overline{p}_{0}-\overline{p}}{\delta \overline{p}}\right),
\label{lambdaexp}
\end{equation}
 where $\delta \overline{p}$ and A are constants, the~latter being easy to obtain by normalization considerations as
\begin{equation}
 A=\frac{1}{\delta \overline{p}}\frac{1}{1-\exp\left(\frac{\overline{p}_{0}-\overline{p}_{L}}{\delta \overline{p}}\right)}.
 \label{lambdaexp2}
\end{equation}

In these equations, $\overline{p}_{0}$ and $\overline{p}_{L}$ are, as~in the previous sections, the~nominal doping at $x=0$ and $x=L$ respectively, being $L$ the size of the film. Note that, by~applying Equation~(\ref{lambda}), this corresponds to:
\begin{equation}\label{exp}
p(x)=\overline{p}_0-\delta \overline{p} \ln \left\{  1- \frac{x}{L} \left[ 1-{\exp}\left(  \frac{\overline{p}_0-\overline{p}_L}{\delta \overline{p}} \right) \right]  \right\}.
\end{equation}

For the case of YBCO films considered in this article, and~for $\overline{p}_{0}=0.135$ and $\overline{p}_{L}=0.161$ as in the previous section, we found that the $\delta \overline{p}$ value that best optimizes the bolometric characteristics (most~notably $\Delta T$) is $\delta \overline{p}=0.007$. We also employed in our evaluations the same common parameter values as described in Sections~\ref{rounding} to~\ref{FEl}.

In the upper row of Figure~\ref{fig5}, the~corresponding doping profile is pictured, both as a $\overline{p}(x)$ representation and as a 2D color density plot. It may be noticed that at the qualitative level the $\overline{p}(x)$ function itself is not too dissimilar to an exponential (however a purely exponential dependence of $\overline{p}$ with $x$ would produce less optimized bolometric performance).

\subsection{Results for the R(T) Profile and Operational Parameters for Resistive TES Use}

\begin{figure}[!ht]
\centering
\includegraphics[]{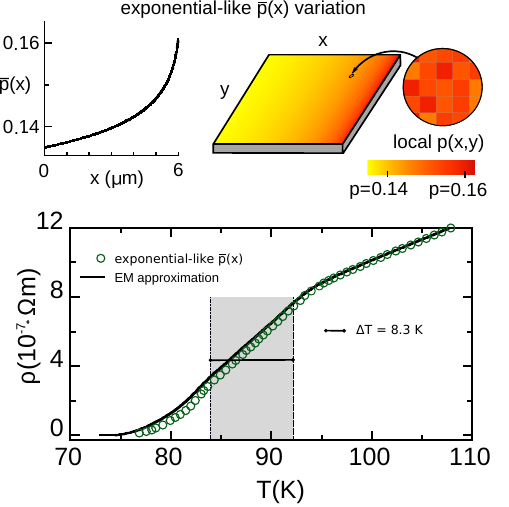}
\caption{In the upper row, we illustrate a YBCO film patterned following the exponential-like variation of nominal doping $\overline{p}(x)$ studied by us in Section~\ref{exponential} (see Equations~(\ref{lambdaexp}) to~(\ref{exp})). We also illustrate doping across the film as a 2D color map, taking into account that the local doping level $p(x,y)$ (zoom in the picture) results from accumulating the exponential-like $\overline{p}(x)$ and the random Gaussian disorder at smaller length scales of about (30 nm)$^2$ (see main text for details). In~the lower row, we plot the resistivity vs. temperature $\rho(T)$ that we obtain for such film (data points for finite-element computations, continuous line for the analytical EM approximation, adapted in this work to this $\overline{p}(x)$ case, see Equation~(\ref{RdeTME})). }
\label{fig5}
\end{figure}

The results of our numerical finite-element evaluation for this $\overline{p}$-pattern are displayed in the second raw of Figure~$\ref{fig5}$ (see also Table~\ref{table1}). As~evidenced in that Figure~\ref{fig5}, not only the transition is significantly broadened with respect to nonpatterned HTS but also (unlike what happened in the case of a linear $\overline{p}(x)$ variation) $\Delta T$ is highly increased. In~particular, the~$\Delta T$ region is increased up to $8.3$ K, almost 10 times more than for nonstructured~HTS.

As may be seen in Table~\ref{table1}, the improvements also occur in the $P^{\rm max}$ parameter, that increase about one order of magnitude with respect to nonstructured HTS. However, note that the TCR value is one order of magnitude worse than in the case of such conventional, nonstructured HTS. This shortcoming and other improvements will be addressed in Section~\ref{discrete} with an evolved $\overline{p}$-pattern~design.

\subsection{Verification Using the Extended-EM Analytical~Approximation}

We have checked our results also using an analytical estimate, by~adapting to this pattern the effective-medium approach described in Section~\ref{EMpat}. For~that, we have combined Equation~(\ref{Rlam}) with the Equation~(\ref{exp}) defining this $\overline{p}$-pattern, to~obtain the new formula:
\begin{equation}
R(T)= \frac{AL}{S} \int_{\overline{p}_0}^{\overline{p}_L} \exp\left(\frac{\overline{p}_{0}-\overline{p}}{\delta \overline{p}}\right)  \frac{{ d\overline{p}} }{\sigma({ \overline{p}},T)}  \ \ \ \mbox{(exp-like pattern)},
 \label{RdeTME}
 \end{equation}
where $\sigma(\overline{p},T)$ results from the Equation~(\ref{inhom}). 
The result of this analytical estimate is displayed in Figure~\ref{fig5} as a continuous line. Again it slightly overestimates $R(T)$ in the tail of the transition, but~basically confirms the finite-element results. As~already mentioned, the~overestimation is expected to be linked to precursor percolation~effects.

\section{Results for Structured HTS Materials Patterned with a Four-Step Exponential-Like Doping~Variation}\label{discrete}

While the $\overline{p}(x)$-pattern design considered in the previous section produced notable improvements of the bolometric features, at~least two concerns may be expressed about it: First, any current structuration experimental technique~\cite{Villegas,Katzer,Johansen,BendingONE, BendingTWO, BendingTHREE, BendingFOUR,networksONE,networksTWO,Verde} may have difficulties producing such a smooth and exponential-like variation of $\overline{p}$ with $x$ (Equations~(\ref{lambdaexp}) to ~(\ref{exp})). Instead, it would be preferable a simpler and, mainly, {\it discrete}-pattern, \ie, one comprised of a few zones, each with a single $\overline{p}$. This would ease fabrication, \eg, by~means of several stages of deoxygenation of YBCO films using different cover masks in each stage. Secondly, the~continuous-pattern of the previous section presents somewhat worsened TCR value with respect to some of the nonpatterned resistive~HTS.

To address both issues, we consider now the discrete $\overline{p}(x)$ pattern described in the upper row of Figure~\ref{fig6}. This pattern defines four zones, each with a single uniform $\overline{p}$ chosen to optimize the linear region of the transition. These values of $\overline{p}$ follow a discretized version of the exponential pattern:
\begin{equation}
L_i=B \exp \left( \frac{\overline{p}_0-\overline{p_{\rm i}}}{\delta \overline{p}}  \right),
\label{disc}
\end{equation}
where $L_i$ is the length of the zone of nominal doping $\overline{p}_{\rm i}$, and~$B$ is a constant so that $\sum_i L_i=L$. 

We tested the bolometric performance for various doping levels $\overline{p}_i$ of the four zones. We obtained the best results with the set $\overline{p}_i= \{0.136$, $0.141$, $0.145,$ $0.160\}$, that we describe~next.

\subsection{Results for the R(T) Profile and Operational Parameters for Resistive TES Use}

The results for our numerical finite-element evaluation for this $\overline{p}$-pattern are displayed in Figure~\ref{fig6} (see also Table~\ref{table1}). As~evidenced there, the~$R(T)$ transition becomes significantly broad and linear, with~such linear region conveniently starting at $T^{-}=76.6$ K (so that the HTS TES could be operated with the simplest liquid-nitrogen bath, at~77 K). The~corresponding $\Delta T$ is now almost 13~K, the~largest obtained in this paper. Also the TCR value $>5~$K$^{-1}$ is the largest obtained in this work, being almost double than for conventional nonstructured (\ie, maximum $T_c$) YBCO. The~$P^{\rm max}$ values (see Table~\ref{table1}) also reflect these improvements, being again the best among the HTS options considered in this work and more than one order of magnitude larger than for the nonstructured~HTS.

To sum up, these finite-element computations reveal that this relatively simple-to-fabricate $\overline{p}$-pattern produces order-of-magnitude improvements over nonstructured HTS materials in $\Delta T$ and $P^{\rm max}$, and~also a 66\% improvement in~TCR. 

\subsection{Verification Using the Extended-EM Analytical~Approximation}

We have checked our results for this four-step $\overline{p}$(x) pattern also using an analytical estimate, by~adapting to this pattern the approach described in Section~\ref{EMpat}. In~this case we used their discretized version , given by Equations~(\ref{Rde1}). By~combining it with the $\overline{p}(x)$-pattern given by Equation~(\ref{disc}) we now obtain:
\begin{equation}
 R=\frac{B}{S}\sum_{i=1}^{N}\frac{1}{\sigma(\overline{p}_i,T)}\exp\left(\frac{\overline{p}_{0}-\overline{p}_i}{\delta \overline{p}}\right),
 \label{RdeTMEdiscrete}
\end{equation}
where the $\overline{p}_i $ are the nominal dopings of each of the $i$ zones, and~$\sigma(\overline{p}_i,T)$ results from Equation~(\ref{inhom}). This analytical estimate is displayed in Figure~\ref{fig6} as a continuous line. It fully confirms the main features obtained by the finite-element computations. Similarly to the case of the other $\overline{p}$-patterns considered in our work, the~estimate is expected to be less reliable in the lower part of the~transition.

\section{Conclusions}\label{conclusion}
To sum up, we considered the advantages of structuring and patterning of the doping level (and~hence of the critical temperature) in high-temperature superconductors with respect to their operational characteristics as resistive bolometric sensors (resistive HTS TES) of electromagnetic radiation. {In particular we studied some chosen examples of spatial variations of the carrier doping into the CuO$_2$ superconducting layers due to oxygen off-stoichiometry.} Our main results are (see also Table~\ref{table1} for a quantitative account):

{\it {(i)}} Non-patterned structured HTS materials (\ie, those with a nominal doping level uniform in space but that does not maximize the critical temperature, thus having random $T_c$-nanostructuring) may already provide some benefit for bolometric use with respect to the nonstructured HTS materials used up to now for those devices. In~particular, they present a widened transition leading to a larger operational temperature interval $\Delta T$ and also larger $P^{\rm max}$ (corresponding to the larger maximum detectable radiation power before sensor saturation). However, these improvements come at the expense of a certain reduction of the sensibility of the sensor as measured by the TCR~value.

\begin{figure}[!ht]
\centering
\includegraphics[]{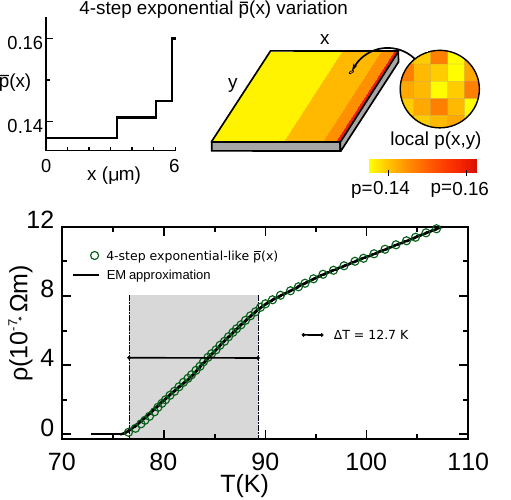}
\caption{In the upper row, we illustrate a YBCO film patterned following the discretized four-step exponential-like variation of the nominal doping $\overline{p}(x)$ studied by us in Section~\ref{discrete}. We also illustrate doping across the film as a 2D color map, taking into account that the local doping level $p(x,y)$ (zoom in the picture) results from accumulating the four-step exponential $\overline{p}(x)$ and the random Gaussian disorder at smaller length scales of about (30 nm)$^2$ (see main text for details). In~the lower row, we plot the resistivity vs. temperature $\rho(T)$ that we obtain for such film (data points for finite-element computations, continuous line for analytical EM approximation, see Equation~(\ref{RdeTMEdiscrete})). This is the most optimized design found in this work for the use in resistive HTS TES devices (see also Table~\ref{table1}).}
\label{fig6}
\end{figure}

{\it (ii)} The bolometric performance may be significantly more optimized with the use of HTS materials including an additional regular dependence on the position of the nominal doping level, $\overline{p}(x)$ (doping-level patterning). In~that case, ad-hoc pattern designs may be found by progressively seeking widened and linear $R(T)$ transitions. Our more optimized design is shown in Figure~\ref{fig6} and consists of just four zones of different sizes and doping levels (related by the exponential-like Equation~(\ref{disc}) evaluated at $\overline{p}_i=0.136$, $0.141$, $0.145$, and $0.160$). With~this design the operational temperature is conveniently located at $T^{-}=76.6$ K, the~operational temperature interval $\Delta T$ is almost 13 K (more than one order of magnitude larger than for the conventional nonstructured YBa$_2$Cu$_3$O$_{6.93}$), the~TCR value is $>5$~K$^{-1}$ (almost double than for the nonstructured case), and~the $P^{\rm max}$ values are also optimized about one order of~magnitude.


\vspace{6pt}

{\bf Acknowledgments:} {This work was supported by projects FIS2016-79109-P (AEI/ FEDER, UE) and AYA2016-78773-C2-2-P(AEI/FEDER,UE), by~the Xunta de Galicia  under grants ED431D 2017/06 and ED431C 2018/11, the Conseller\'{\i}a de Educaci\'on Program for Development of a Strategic Grouping in Materials AeMAT under Grant No. ED431 {2018/08}, Xunta de Galicia,  and by the CA16218 nanocohybri COST Action. JCV thanks the Spanish Ministry of Education for grant FPU14/00838.}

\mbox{}

{\bf  Abbreviations:} The following abbreviations are used in this manuscript:\\\mbox{}\vspace{0.1cm}\\
\begin{tabular}{@{}ll}
TCR & Temperature coefficient resistance\\
YBCO & YBa$_2$Cu$_3$O$_{\delta}$\\
STO & SrTiO$_3$ \\
CMOS & silicon/YSZ/Zirconia \\
{YSZ} & {Yttria-stabilized zirconia}\\
TES & Transition-edge sensor\\
HTS & High-temperature superconductor \\
FWHM & Full width at half maximum \\
EM & Effective medium \\
BKT & Berezinskii--Kosterlitz--Thouless
\end{tabular}

\begin{table}[!ht]
\caption{Summary of the main operation parameters for resistive high-$T_c$ cuprate superconductors (HTS) transition-edge bolometer devices (TES) devices using, as sensing materials, YBCO superconductors with the various doping nanostructurations and patterns explored in this work. These include the usual nonstructured  YBCO ($\overline{p}=0.155$) and  the novel options considered by us: structured nonpatterned YBCO ($\overline{p}\leq 0.155$, Section~\ref{section4}) and  structured patterned YBCO (Sections~\ref{constant} to~\ref{discrete}). The~$\Delta T$ and TCR follow from the $\rho(T)$ curves presented in Figures~\ref{figA} and~\ref{fig3}--\ref{fig6}. The~$P^{\rm max}$ values were calculated for the three example device implementations described in Sections~\ref{microsensor} and~\ref{milisensor} (microbolometers with direct cooling and SrTiO$_3$(STO) or silicon/Yttria-stabilized zirconia (YSZ)/Zirconia (CMOS)-type substrates, plus~millimeter-wave sensors using a meander geometry).}

\mbox{}

\mbox{}

\centering  \footnotesize
 \begin{tabular}{l|cc|ccc}

\hline &  & &  &  & \\

\hfill \textbf{Sensing Material} \hfill\mbox{}
& \textbf{$\mathbf{\Delta T}$}
& \textbf{$\mathbf{TCR}$} 
& {\textbf{$\mathbf{P^{max}}$}} \boldmath$({\upmu \rm W)}$
& {\textbf{$\mathbf{P^{max}}$}} \boldmath$({\upmu \rm W)}$
& {\textbf{$\mathbf{P^{max}}$}} \boldmath$({\upmu \rm W)}$

\\
\hfill \textbf{} \hfill\mbox{}
& \hfill {\boldmath${ (\rm K)}$} \hfill
& \hfill {\boldmath${(\rm K^{-1})}$} \hfill\mbox{}
& \boldmath$\upmu$\textbf{m-sensor}
& \boldmath$\upmu$\textbf{m-sensor}
& \textbf{mm-sensor}

\\
\hfill \textbf{} \hfill\mbox{}
& 
& 
& \textbf{over STO}
& \textbf{over CMOS}
& \textbf{over CMOS}

\\
\hline&  & &  & &  \\

nonstructured YBCO:  \hfill &  & &  & & \\
\hfill    $\overline{p}=0.155$ & 0.9 & 3.05 & $0.5$ & $0.037$ & 13  \\

&  & &  &  &  \\
\hline&  & &  & & \\
structured nonpatterned YBCO:\;\; &  & &  &  & \\

\hfill  $\overline{p}=0.147$ & 1.8 & 2.67 & $1.1$ & $0.080$ & 31\\

\hfill  $\overline{p}=0.140$ & 2.3 & 2.15  &$1.3$ & $0.095$ & 38 \\

\hfill  $\overline{p}=0.135$ & 4.1 & 2.16 & $2.3$ & $0.17$ & 72 \\

&  & &  & &  \\
\hline&  & &  & &  \\
structured patterned YBCO: \hfill &  & &  & & \\

\hfill linear $\overline{p}(x)$ & 1.4 & 0.25 & $0.8$ & $0.062$ & 4.2 \\

\hfill exponential-like $\overline{p}(x)$& 8.3 & 0.18 & $4.9$ & $0.35$ & 120 \\

\hfill 4-step exponential-like $\overline{p}(x)$& 12.7 & 5.13  & $7.2$ & $0.55$ & 230  \\

&  & &  & &  \\
\hline
\end{tabular}

\label{table1}

\end{table}


\clearpage

\begin{thebibliography}{999}


\bibitem{HTSbol9} Mohajeri, R.; Opata, Y.A.; Wulff, A.C.; Grivel, J.C.; Fardmanesh, M. All Metal Organic Deposited High-Tc Superconducting Transition Edge Bolometer on Yttria-Stabilized Zirconia Substrate. {\em J. Supercond. Nov. Magn.} {\bf 2017}, {\em 30}, 1981--1986.

\bibitem{HTSbol4} Moftakharzadeh, A.; Kokabi, A.; Banzet, M.; Schubert, J.; Fardmanesh, M. Detectivity Analysis and Optimization of Large-Area Freestanding-Type HTS Bolometers.  {\em IEEE Tran. Appl. Supercond.} {\bf 2012}, {\em 22}, 2100107.

\bibitem{HTSbol1} Hosseini, M.; Moftakharzadeh, A.; Kokabi, A.; Vesaghi, M.A.; Kinder, H.; Fardmanesh, M. Characterization of a Transition-Edge Bolometer Made of YBCO Thin Films Prepared by nonfluorine Metal-Organic Deposition. {\em IEEE Tran. Appl. Supercond.} {\bf 2011}, {\em 21}, 3587--3591.

\bibitem{HTSbol5} Moftakharzadeh, A.; Kokabi, A.; Ghodselahi, T.; Vesaghi, A.; Khorasani, S.; Banzet, M.; Schubert, J.; Fardmanesh, M. Investigation of Bias Current and Modulation Frequency Dependences of Detectivity of YBCO TES and the Effects of Coating of Cu-C Composite Absorver Layer. {\em IEEE Tran. Appl. Supercond.} {\bf 2009}, {\em 19}, 3688--3693.

\bibitem{HTSbol11} Hosseini, M.; Moftakharzadeh, A.; Kokabi, A.; Vesaghi, M.A.; Kinder, M.; Fardmanesh, M. 2D Analysis of the Effects of Geometry on the Response of High-$T_c$ Superconductive Bolometric Detectors. {\em IEEE Tran. Appl.~Supercond.} {\bf 2009}, 19, 484--488.

\bibitem{Oktem} Oktem, B.; Bozbey, A.; Avci, I.; Tepe, M.; Abukay, D.; Fardmanesh, M. The superconducting transition width and illumination wavelength dependence of the response of MgO substrate YBCO transition edge bolometers. {\em Phys. C} {\bf 2007}, {\em 458}, 6--11.

\bibitem{HTSbol6} Lakew, B.; Brasunas, J.C.; Aslam, S.; Pugel, D.E. High-Tc transition edge superconducting (TES) bolometer on a monolithic saphire membrane-- construction and performance. {\em Sensors Actuators A.} {\bf 2004}, {\em 114}, 36--40.

\bibitem{HTSbol7} Delerue, J.; Gaugue, A.; Testé, P.; Caristan, E.; Klisnick, G.; Redon, M.; Kreisler, A. YBCO Mid-Infrared Bolometer Arrays. {\em IEEE Tran. Appl. Supercond.} {\bf 2003}, {\em 13}, 176--179.



\bibitem{Tkachenko} Khrebtov, I.A.; Tkachenko, A.D. High-Temperature superconductor bolometers for the IR region. {\em \mbox{J. Opt. Technol.}} {\bf 1999}, {\em 66}, 735--741.

\bibitem{Kaiser} Kaiser, G.; Thurk, M.; Seidel, P. Signal-to-noise ratio improvement of HTSC bolometers for cryocooler operation using a thermal compensation principle. {\em Cryogenics} {\bf 1995}, {\em 35}, 463--465.

\bibitem{HTSbol12} Zakar, E.; Wikner, D.; Dubey, M.; Amirtharaj, P. Thin Film YBCO Pixels for MMW Detector. {\em Adv. Sci. Technol. Res. J.} {\bf 2008}, {\em 54}, 434--438.

\bibitem{HTSbol8} Mohajeri, R.; Nazifi, R.; Wulff, A.C.; Vesaghi, M.A.; Grivel, J.C.; Fardmanesh, M. Investigation of CeO$_{2}$ Buffer Layer Effects on the Voltage Response of YBCO Transition Edge Bolometers. {\em IEEE Tran. Appl. Supercond.} {\bf 2016}, {\em 26}, 1--4.

\bibitem{Nivelle} de Nivelle, M.J.M.E.; Bruijn, M.P.; de Vries, R.; Wijnbergen, J.J.; de Korte, P.A.J.; S\'anchez, S.; Elwenspoek, M.; Heidenblut, T.; Schwierzi, B.; Michalke, W.; et al.  Low noise high-$T_c$ superconducting bolometers on silicon nitride membranes for far-infrared detection. {\em J. Appl. Phys.} {\bf 1997}, {\em 82}, 4719--4726.

\bibitem{Frenkel} Zhang, Z.M.; Frenkel, A. Thermal and Nonequilibrium Responses of Superconductors for Radiation Detectors. {\em J. Supercond.} {\bf 1994}, {\em 7}, {871--884}.

\bibitem{Khrebtov} Khrebtov, I.A. Noise Properties of High Temperature Superconducting Bolometers. \textit{Fluct. Noise Lett.} {\bf 2002}, {\em 2}, 51--69.

\bibitem{HTSbol3} Ivanov, K.V.; Khokhlov, D.A.; Khrebtov, I.A.; Kulikov, Y.V.; Malyarov, V.G.; Nikolenko, A.D.; Pindyurin, V.F.; Zerov, V.Y. Characterization of the composite bolometer with a high-Tc superconductor thermometer for an absolute radiometer of synchrotron radiation. {\em Nucl. Instrum. Methods Phys. Res. A.} {\bf 2007}, {\em 575}, 272--275.



\bibitem{Harry}  Eaton, H.A.C. Observing Photons in Space, chapter 29. In {\em Infrared Imaging Bolometers}; Huber, M.C.E., Pauluhn,~A., Culhane, J.L., Timothy, J.G., Wilhelm, K., Zehnder, A., Eds.; AG of the Series ISSI Scientific Report Series; Springer: Basel, Switzerland, {2013}; pp. 515--524, Volume 9.

\bibitem{Irwin}Irwin, K.D. An application of electrothermal feedback for high resolution cryogenic particle detection. {\em \mbox{Appl. Phys. Lett.}} {\bf 1995}, {\em 66}, 1998--2000.

\bibitem{Irwinlibro}  Irwin, K.D.; Hilton, G.C. Transition-Edge Sensors in Cryogenic Particle Detection, edited by C. Enss, Springer-Verlag, Berlin Heidelberg. {\em Top. Appl. Phys.} {\bf 2005}, {\em 99}, 63--149.


\bibitem{Abdel} Abdel-Rahman, M.; Ilahi, S.; Zia, M.F.; Alduraibi, M.; Debbar, N.; Yacoubi, N.; Ilahi, B. Temperature coefficient of resistance and thermal conductivity of Vanadium oxide 'Big Mac' sandwich structure. {\em \mbox{Infrared Phys. Technol.}} \textbf{2015}, \emph{71}, 127--130.


\bibitem{Villegas} Crassous, A.; Bernard, R.; Fusil, S.; Bouzehouane, K.; Bourdais, D.L.; Enouz-Vedrenne, S.; Briatico, J.; Bibes, M.; Barthélémy, A.; Villegas, J.E. Nanoscale electrostatic manipulation of magnetic flux quanta in ferroelectric/superconductor BiFeO$_3$/YBa$_2$Cu$_3$O$_{7-\delta}$ heterostructures. {\em Phys. Rev. Lett.} \textbf{2011}, \emph{107}, 247002.

\bibitem{Katzer} Katzer, C.; Stahl, C.; Michalowski, P.; Treiber, S.; Schmidl, F.; Seidel, P.; Albrecht, J.; Schütz, G. Gold nanocrystals in high-temperature superconducting films: Creation of pinning patterns of choice. {\em New J. Phys.} \textbf{2013}, \emph{15}, 113029.

\bibitem{Johansen} Vestg{\aa}rden, J.I.; Yurchenko, V.V.; W\"ordenweber, R.;  Johansen, T.H. Mechanism for flux guidance by micrometric antidot arrays in superconducting films. {\em Phys. Rev. B} \textbf{2012}, \emph{85}, 014516.



\bibitem{BendingONE}Moschalkov, V.V. Nanostructured superconductors: quantum matter at low temperatures.  In Proceedings of the Workshop on vortex behavior in unconventional superconductors, Braga, Portugal, 7--12 October 2018.

\bibitem{BendingTWO}Feuiller-Palma, C. High-Tc superconducting devices. In Proceedings of the Coherent Superconducting Hybrids and Related Materials, Les Arcs 1800, France, 26--29 March 2018.

\bibitem{BendingTHREE}Garc\'{\i}a-Serrano, I.;  C\'ordoba, R.;  Ses\'e, J.;  Ibarra, M.R.;  Guillam\'on, I.;  Suderow, H.;  Vieira, S.;  De Teresa, J.M. Superconducting nanostructures grown by focused ion beam induced deposition. In Proceedings of the International conference on nano confined superconductors and their application, Garmish, Germany, 3--7 September 2016.

\bibitem{BendingFOUR}Ramallo, M.V.;  Carballeira, C.;  Cot\'on, N.;  Mosqueira, J.;  Ramos-\'Alvarez, A.; Vidal, F. Influence of disorder and reduced dimensionality on the critical phenomena around the superconducting transition in cuprates. In Proceedings of the International workshop on advances in nanostructured superconductors: materials, properties and theory, Madrid, Spain, 4--7 May 2014.



\bibitem{networksONE} European network NanoSC. Available online: {\sf http://www.cost.eu/actions/MP1201}  ({accessed on 1 January 2020}).  

\bibitem{networksTWO} European network NanoCoHybri. Available online: {\sf http://www.cost.eu/actions/CA16218} ({accessed on 1 January 2020}).

\bibitem{Verde} Verde, J.C.; Doval, J.M.; Ramos-Álvarez, A.; Sóñora, D.; Ramallo, M.V. Resistive Transition of High-$T_c$ Superconducting Films With Regular Arrays of $T_c$-Domains Induced by Micro- or Nanofunctionalization. {\em IEEE Tran. Appl. Supercond.} \textbf{2016}, \emph{26}, 8800204.

\bibitem{y} Maza, J.; Vidal, F. Critical-temperature inhomogeneities and resistivity rounding in copper oxide superconductors. { \em Phys. Rev. B.} {\bf 1991}, {\em 43}, 10560. 



\bibitem{Coton2} Cot\'on, N; Mercey, B.; Mosqueira, J.; Ramallo, M.V.; Vidal, F. Synthesis from separate oxide targets of high quality $La_{2-x}Sr_xCuO_4$ thin films and dependence with doping of their superconducting transition width. {\em Supercond. Sci. Technol.} {\bf 2013}, 26, 075011.


\bibitem{Coton3} Cot\'on, N.;  Ramallo, M.V.; Vidal, F. Critical temperatures for superconducting phase-coherence and condensation in $La_{2-x}Sr_xCuO_4$. {\em arXiv} {\bf 2013}, arXiv:1309.5910v3.

\bibitem{a-benfatto} Benfatto, L.; Castellani, C.; Giamarchi, T. Doping dependence of the vortex-core energy in bilayer films of cuprates. {\em Phys. Rev. B} {\bf 2008}, {\em 77}, 100506.

\bibitem{b-benfatto} Caprara, S.; Grilli, M.; Benfatto, L.; Castellani, C. Effective medium theory for superconducting layers: A~systematic analysis including space correlation effects. {\em Phys. Rev. B} {\bf 2011}, {\em 84}, 014514.


\bibitem{kappa}  Yu, C.; Scullin, M.L.; Ramamoorthy, M.H.; Majumdar, A.  Thermal conductivity reduction in oxygen-deficient strontium titanates. {\em Appl. Phys. Lett.} {\bf 2008}, {\em 92}, 191911

\bibitem{Puica} Lang, W.; Puica, I.; Zechner, G.; Kitzler, T.; Bodea, M.A.; Siraj, K.; Pedaring, J.D. All Non-ohmic electrical transport properties above the critical temperature in optimally and underdoped superconducting $YBa_{2}Cu_{3}O_{6+x}$. {\em J. Supercond. Nov. Magn.} {\bf 2012}, {\em 25}, 1361--1364.



\bibitem{PNAS} Barišić, M.; Chan, M.K.; Li, Y.; Yu, G.; Zhao, X.; Dressel, M.; Smontara, A.; Greven, M. Universal sheet resistance and revised phase diagram of the cuprate high-temperature superconductors. {\em Proc. Natl. Acad. Sci. USA} \textbf{2013}, \emph{110}, 12235--12240.

\bibitem{Nature3} Barišić, N.;  Badoux, S.; Chan, M.K.; Dorow, C.; Tabis, W.; Vignolle, B.; Yu, G.; Béard, J.; Zhao, X.; Proust,~C.; et~al. Universal quantum oscillations in the underdoped cuprate superconductors. {\em Nat. Phys.} {\bf 2013}, {\em 9}, 761--764. 

\bibitem{Tafti} Tafti, F.F.; Lalibert\'e, F.; Dion, M.; Gaudet, J.; Fournier, P.; Taifeller, L. Nernst effect in the electron-doped cuprate superconductor $Pr_{2-x}Ce_xCuO_4$: Superconducting fluctuations, upper critical field $H_{c2}$, and the origin of the $T_c$ dome. {\em Phys. Rev. B} {\bf 2014}, {\em 90}, 024519.

\bibitem{a} Pomar, A.; Ramallo, M.V.; Maza, J.; Vidal F. Measurements of the fluctuation-induced magnetoconductivity in the $a$-direction of an untwinned Y$_1$Ba$_2$Cu$_3$O$_{7-\delta}$ single-crystal in the weak magnetic field limit. {\em Phys. C} {\bf 1994}, {\em 225}, 287--293.

\bibitem{c} Ramallo, M.V.; Vidal, F. On the width of the full-critical region for thermal fluctuations around the superconducting transition in layered superconductors. {\em Europhys. Lett.} {\bf 1997}, {\em 39}, 177--182.

\bibitem{b} Vi\~na, J.; Camp\'a, J.A.; Carballeira, C.; Curr\'as, S.R.; Maignan, A.; Ramallo, M.V.; Rasines, I.; Veira, J.A.; Wagner, P.; Vidal, F. Universal behavior of the in-plane paraconductivity of cuprate superconductors in the short-wavelength fluctuation regime. {\em Phys. Rev. B} {\bf 2002}, {\em 65}, 212509-1--212509-4.

\bibitem{Carballeira} Carballeira, C.; Currás, S.R.; Viña, J.; Veira, J.A.; Ramallo, M.V.; Vidal, F. Paraconductivity at high reduced temperatures in $YBa_2Cu_3O_{7-\delta}$ superconductors. {\em Phys. Rev. B} {\bf 2001}, {\em 63}, 144515.

\bibitem{Ramallo} Ramallo, M.V.; Pomar, A.; Vidal, F. In-plane paraconductivity and fluctuation-induced magnetoconductivity in biperiodic layered superconductors: Application to $YBa_2Cu_3O_{7-\delta}$. {\em Phys. Rev. B.} {\bf 1996}, {\em 54}, 4341--4356.

\bibitem{Mosqueira} Mosqueira, J.; Recolevschi, A.; Torr\'on, C.; Ramallo, M.V.; Vidal, F. Crossing point of the magnetization versus temperature curves and the Meissner fraction in granular $La_{1.9} Sr_{0.1}CuO_4$ superconductors: Random orientation and inhomogeneity effects. {\em Phys. Rev. B} {\bf 1999}, {\em 59}, 4394--4403.

\bibitem{Coton} Cot\'on, N.; Ramallo, M.V.; Vidal, F. Effects of critical temperature inhomogeneities on the voltage-current characteristics of a planar superconductor near the Berezinskii-Kosterlitz-Thouless transition. {\em \mbox{Supercond. Sci. Technol.}} {\bf 2011}, {\em 24}, 085013.



\bibitem{Solov} Solov'ev, A.L.; Dmitriev, V.M. Fluctuation conductivity and pseudogap in YBCO high-temperature superconductors (Review). {\em J. Low Temp. Phys.} {\bf 2009}, {\em 35}, 169--197.

\bibitem{LD} Lawrence, W.E.; Doniach, S. {Theory of layer structure superconductors}. In Proceedings of the Twelfth International Conference on Low-Temperature Physics, Kyoto, Japan, 4--10  September 1970;  Kanda, E., Ed.; ~Keigatu: Tokyo, Japan, 1971; p. 361.


\bibitem{d} Carballeira, C.; Mosqueira, J.; Ramallo, M.V.; Veira, J.A.; Vidal, F. Fluctuation-induced diamagnetism in bulk isotropic superconductors at high reduced temperatures. {\em J. Phys. Condens. Matter} {\bf 2001}, {\em 13}, 9271--9279.

\bibitem{e} Mosqueira, J.; Ramallo, M.V.; Curr\'as, S.R.; Torr\'on, C.; Vidal, F. Fluctuation-induced diamagnetism above the superconducting transition in MgB$_2$. {\em Phys. Rev. B} {\bf 2002}, {\em 65}, 174522-1--174522-7.


\bibitem{Ying} Ying, Q.Y.; Kwook, H.S. Kosterlitz-Thouless transition and conductivity fluctuations in Y-Ba-Cu-O thin films. {\em Phys. Rev. B} {\bf 1990}, {\em 42}, 2242--2247.



\bibitem{Halperin} Halperin, B.I.; Nelson, D.R. Resistive Transition in Superconducting Films. {\em J. Low Temp. Phys.} {\bf 1979}, {\em 36}, 599--616.

\bibitem{Mosqueira2} Mosqueira, J.; Cabo, L.; Vidal, F. Structural and $T_c$ inhomogeneities inherent to doping in $La_{2-x}Sr_{x}CuO_4$ superconductors and their effects on the precursor diamagnetism. {\em Phys. Rev. B} {\bf 2009}, \emph{80}, 214527.

\bibitem{Mihailovic} Mihailovic, D.; Optical experimental evidence for a universal length scale for the dynamic charge inhomogeneityy of cuprate superconductors. {\em Phys. Rev. Lett.} {\bf 2005}, {\em 94}, 201001.

\bibitem{lbts} Verde, J.C.; Ramallo, M.V. Herramientas Computacionales en el laboratorio LBTS. In {\em Pel\'iculas Micro- y Nanoestructuradas de Superconductores de alta Temperatura: Computaci\'on de su Transici\'on Resistiva;} GIDFI: Santiago de Compostela, Spain, 2016;  pp. 39--47, ISBN 978-1-68073-062-3.

\bibitem{x} Garland, J.C.; Tanner, B.D. (Eds.) \emph{Electrical Transport and Optical Properties of Inhomogeneous Media}; AIP: New~York, NY, USA, 1978; p. 2.

\bibitem{Tallon} Tallon, J.L.; Bernhard, C.; Shaked, H. Generic superconducting phase behavior in high-$T_c$ cuprates: $T_c$ variation with hole concentration in $YBaCu_3O_{7-\delta}$. {\em Phys. Rev. B.} {\bf 1995}, {\em 51},  12911--12914.







\end{thebibliography}
\end{document}